\documentclass[runningheads]{llncs}
\usepackage{graphicx}
\usepackage{multirow}
\usepackage[table]{xcolor}
\usepackage[belowskip=-15pt,aboveskip=5pt, font=small, labelfont=bf]{caption}
\begin{document}
\title{Stress Management Using Virtual Reality-Based Attention Training
\thanks{This study has been approved by the IRB Approval Committee at VRapeutic Inc. and the Research Ethics Approval Committee of the Electronics and Communications Engineering Department at the Arab Academy for Science, Technology, and Maritime Transport.}}
\author{Rojaina Mahmoud\inst{1}\orcidID{0009-0009-4249-3310}\and
Mona Mamdouh\inst{1}\orcidID{0009-0008-2071-2095}\and
Omneya Attallah\inst{1}\orcidID{0000-0002-2657-2264} \and Ahmad Al-Kabbany \inst{1,2} \orcidID{0000-0002-3941-0061}}
\authorrunning{Attallah et al.}
\institute{Electronics and Communications Engineering Department, Arab Academy for Science, Technology, and Maritime Transport, Alexandria 21937, Egypt
\email{o.attallah@aast.edu,alkabbany@ieee.org}\\
%Intelligent Systems Lab, Arab Academy for Science, Technology, and Maritime Transport, Alexandria 21937, Egypt
%\email{alkabbany@ieee.org,alkabbany@aast.edu}\\
%\url{https://www.linkedin.com/company/intelligent-systems-lab} 
\and
VRapeutic Inc., Ottawa K2M1T2, Canada}

\maketitle              % typeset the header of the contribution
\begin{abstract}
In this research, we are concerned with the applicability of virtual reality-based attention training as a tool for stress management. Mental stress is a worldwide challenge that is still far from being fully managed. This has maintained a remarkable research attention on developing and validating tools for detecting and managing stress. Technology-based tools have been at the heart of these endeavors, including virtual reality (VR) technology. Nevertheless, the potential of VR lies, to a large part, in the nature of the content being consumed through such technology. In this study, we investigate the impact of a special type of content, namely, attention training, on the feasibility of using VR for stress management. On a group of fourteen undergraduate engineering students, we conducted a study in which the participants got exposed twice to a stress inducer while their EEG signals were being recorded. The first iteration involved VR-based attention training before starting the stress task while the second time did not. Using multiple features and various machine learning models, we show that VR-based attention training has consistently resulted in reducing the number of recognized stress instances in the recorded EEG signals. This research gives preliminary insights on adopting VR-based attention training for managing stress, and future studies are required to replicate the results in larger samples.\footnote{All the authors contributed equally to this research}

\end{abstract}

\section{Introduction}
Mental stress is one of the most significant health threats worldwide. While it is not getting less significant due to ongoing global social and economic factors, the continuous increase of stress levels can put the sustainability of a society’s productivity at stake. This has maintained a remarkable research attention on developing and validating various tools for stress detection and management. A significant share of these tools is for emerging technologies. To a large part, this stems from the rise of pervasive and mobile computing, including wearables and immersive technologies. Virtual reality (VR) as one sub-category of immersive technologies, with the power of full immersion being its core premise, has shown a growing potential as a framework for stress management.

Not just in stress management, VR has been gaining momentum in several areas under the umbrella of mental health and neuromodulation for decades. The most recent hardware advancements though have facilitated an unprecedented pace for incorporating VR in the healthcare sector. This is apparent in the growing literature on VR for pain management using active distraction, psychological disorders, motor disorders in children and adults, and cognitive disorders including attention deficit. A typical VR-based attention training involves customizable scenarios in which the trainee is asked to accomplish a certain task that requires one or more of the various attention types, such as sustained, selective, and adaptive attention. This usually takes place while performance metrics are being monitored and recorded/computed from within the virtual environment. These metrics include but are not limited to response time, impulsivity, and omission score.

The research literature on attention training has featured multiple frameworks, some of which are non-technology related. For the sake of relevance, we give technology-based approaches more attention in this research. Several studies have addressed the feasibility of attention training using various forms of technology. These include mobile and virtual reality applications. Due to the variety of adopted frameworks, some researchers are interested in identifying the variables that influence the outcomes of an attention training, which is expected to facilitate taking informed system design decision. Another notable research direction in the literature on attention training has investigated its influence together with that of cognitive modification on other disorders including anxiety and depressive disorders. Attention restored through mindfulness was also shown to increase one’s capacity to manage stress. This line of research is the principal inspiration for this study.

In this research, we are concerned with the feasibility of adopting a technology-based attention training, particularly using VR, to help people manage stress. Throughout this study, the mental stress state is induced using a mental arithmetic task, i.e., MAE stressor. We start by training different machine learning (ML) models on binary stress detection, then we used these models to identify stress instances with and without attention training. On a group of fourteen undergraduate engineering students, we exposed them to attention training followed by a stress task, then we expose them a stress task without a preceding attention training. Their EEG signals were recorded in both stages and signal processing techniques, such as pre-filtering and feature extraction, were applied on the collected signals before we fed them into the pre-trained ML models to identify the stress states. With consistency, we report less stress states following an attention training. The contributions of this research can be summarized as follows:
\begin{enumerate}
    \item i.	We constructed a dataset for binary stress detection, i.e., two levels only, either stress or non-stress, from a total of 10 students. The MAE-induced stress was used, and the models were trained on the EEG signals of the volunteering participants. Those participants do not have to be the same participants with which we continue the subsequent stages of the study.
    
    \item We pre-processed the collected data, then we trained three machine learning models, namely, SVM, KNN, and LDA, using wavelet and Modified Frequency Mean (MFMD) features on binary stress detection.
    
    \item We constructed another dataset for testing the trained models where the participants were exposed to two iterations of a stress task with attention training as a preceding step then without attention training. The same type of signal processing operations (like those used in step ii) were used with the testing signals.
    
    \item We show that across different ML models, there is a consistent decline in the number of instances identified as stress samples following attention training (AT).
\end{enumerate}
Given the limited number of participants, this research is meant to give preliminary insights on the potential of VR-based training as a stress management tool.

The rest of this article is structured as follows. In section 2, we highlight the different research direction that are relevant to the proposed study. In section 3, we present the pipeline of the study in detail. Section 4 features the obtained results and discusses their interpretation, before the study is concluded in section 5.

\section{Related Work}
Mental stress is among the most widespread individual afflictions, and it has a substantial influence on our current societies. Many research investigations inspected the adverse impacts of stress on people's health \cite{turner2020psychological,o2021stress}. The need for reliable and suitable stress identification, evaluation, and handling systems that effectively aid humans in identifying their degree of stress and enable them to effectively handle and fight it is expanding. Recent research has concentrated on the early identification of stress as a means to avoid serious medical conditions. Numerous approaches have been employed as stress indicators including Electrocardiogram (ECG), Electromyogram (EMG), and Electroencephalogram (EEG). However, the literature has demonstrated the superiority of EEG among others in detecting stress \cite{petrantonakis2011novel,panicker2019survey}. Multiple investigations have created precise machine learning stress-level classifiers using EEG-based attributes \cite{attallah2020effective,alshorman2022frontal,salankar2022eeg,saputra2022identification,suryawanshi2023brain,kit2022discrete,kotkar2022analysis,pathak2022human,mazlan2023investigation,malviya2022novel}. Nevertheless, these studies focused only on identifying stress and did not consider handling and reducing stress.

Virtual reality (VR) is an immersive technology example that produces a computer-generated setting with realism-based visuals, sounds, and various other emotions. In order to conduct investigations or deliver therapy, numerous scientists have employed VR headgear to create a variety of stimuli. VR is quickly becoming a fascinating and promising useful method for improving an individual's well-being and health \cite{riva2016transforming,pallavicini2019assessing}. This is brought on by its long-term cost-effectiveness and expanding reach. However, regardless of the fact that numerous comprehensive investigations indicate the beneficial effects of this kind of device for mental duties \cite{parsons2017virtual,bouchard2019applications,bouchard2019applications}, less attention seems to be paid to the therapeutic utilization of VR headgear for stress/ emotions evaluation and control \cite{pallavicini2019assessing}. For example, the study \cite{perez2021quantitative} employed a VR relaxation experience to manage stress by exposing the participants to a relaxing video showing natural scenes while measuring their EEG signals. Likewise, the researchers of \cite{adhyaru2022virtual} employed VR relaxation video displaying natural environments to reduce stress of clinicians in their workplaces. Also, in the study \cite{beverly2022tranquil}, in order to reduce  stress for healthcare professionals who work in COVID-19 treatment facilities, a Tranquil Cinematic-VR modeling of a natural setting was implemented.  Furthermore, the research article \cite{bjorling2022exploring} examined the impact of natural scene-based VR video on stress relief of adolescents.  Relaxing VR based on natural scenes is valuable because it has been shown to reduce overall arousal and has favorable impacts on mental stress. However, in real life, such an emotional condition and its positive aspects are rarely sustained for a long time, and the VR experience is scarcely transferable to different contexts \cite{pizzoli2019user}.

On the other hand, other engagement-based VR activities are favorable as they involve the participant in virtual reality-mediated endeavors to strengthen his or her own capacity for controlling emotions. This type of strategy necessitates individuals to engage with virtual content, allowing the development of particular skills, instead of simply implying a passive visualization of the virtual setting or interaction with calming stimuli. This is what happens for certain therapeutic interventions and emotional management training in VR. An example of such engagement activities that could be induced in a VR environment is attention training (AT). AT involves practicing a specific attention control exercise. According to the reviews \cite{fergus2016attention,calkins2015effects},  AT method demonstrated evidence of verification for various kinds of anxiety and signs of depression. Additionally, there is mounting proof that the influence is discernible in neuro-cognitive tests linked to actual attention management \cite{knowles2016systematic}. Furthermore, it has been proven that symptoms can be lessened by using AT, which aims to strengthen flexible, voluntary extrinsic attention.

In this study, we take advantage of VR by creating attention-training exercises in specific virtual settings, and we subsequently analyze how this VR-based AT affects stress management and evaluation. More specifically, we exposed two groups of participants to the commonly utilized stressor known as the MAE, one of which has undergone AT prior to exposure to MAE, and then we compare the two groups' performance using the EEG signals they collected during MAE.

\section{Proposed Method}
In this section, we present, discuss, and justify our key design decision for the two main stages in this study, namely, the electroencephalogram (EEG) signal acquisition and EEG signal analysis.

\subsection{Experimental Procedure}
\subsubsection{Training Phase}
\begin{figure}
\begin{center}
\setlength{\belowcaptionskip}{-30pt plus 0pt minus 0pt}
\includegraphics[width=0.5\textwidth]{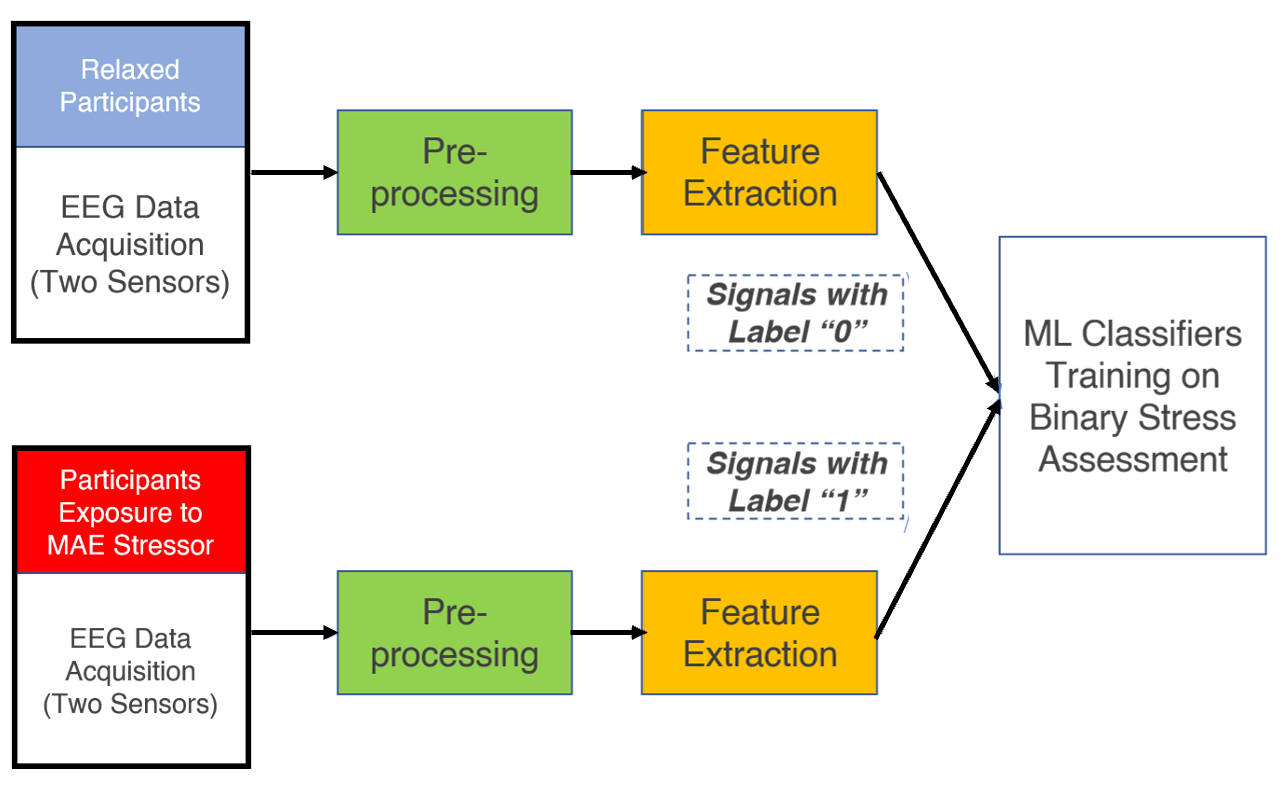}
\caption{The training pipeline adopted in this research. Please see text for more details.} \label{fig:train}
\end{center}
\end{figure}

Ten undergraduate engineering students were recruited for the training phase. The same group of students were instructed to go through the two arms shown in Fig.~\ref{fig:train}. The upper part of the figure involved exposing the recruited subjects to a relaxing routine for a duration of four minutes. This is realized by instructing the participants to listen to a standard relaxing piano music. The lower part of the figure is realized by exposing the participants to three minutes of mental arithmetic operations of escalating difficulty. The operations started with adding two numbers then solving a system of linear equations then solving an arithmetic expression that is composed of addition, subtraction, multiplication, and division, with parentheses added to modify precedence. The EEG signals were acquired amidst the two processes of relaxation and exposure-to-stress using Muse S\textsuperscript{TM} EEG headset. X\% of the acquired samples is used for training and the left is left for testing. %More details on the processing and the analysis of the acquired EEG signals will be given in the rest of this section.

\subsubsection{Testing Phase}
Three participants, from the previously recruited ten students in the training phase, were allowed to participate in the testing phase. Following a duration of a two-hour break, they were instructed to receive a VR-based AT for a duration of ninety seconds. This attention training was realized using Viblio\textsuperscript{TM} that is developed by VRapeutic\footnote{http://myvrapeutic.com/}\cite{VRapeutic1}\cite{VRapeutic2}. This VR module features a book ordering task in which the player is required to put-in-order a varying number of books on four book shelves withing a certain amount of time. To be completed, this requires sustained attention for the duration of the task. Following the AT, the participants were exposed to a mental arithmetic task. Similar to the training phase, the EEG signals were acquired amidst the mental arithmetic task. The testing dataset is comprised of two components: 1) stress samples from the dataset acquired during the training phase (with no overlap with the training dataset), and 2) stress samples acquired after the VR-based AT.

\subsection{EEG Signal Processing and Analysis}
Matlab\textsuperscript{TM} was adopted as the main scientific computing tool in this study. We also used the EEG Lab library to realize the basic handling, management, and processing operations of signals. Various operations for pre-processing were implemented using EEG Lab. We commenced by a band-pass filtering stage for the data in the range from 1 to 30 Hz. Then, we applied Independent Component Analysis (ICA) for separating the independent linearly mixed sources in the data. Relevant components were then determined and retained, and the rest are eliminated, using the function \textit{ICLabel} of EEG Lab. %Following extensive simulations, we had found that the re-referencing of data is essential after each pre-processing step. This re-referencing was followed in all of our simulations. Moreover, during both training an testing, we carried out windowing to all the data under consideration. 
Lastly, wavelet denoising with four levels was applied on the data before extracting two types of features, namely, Modified Frequency Mean (MFMD) features and wavelet features. For the latter, four-level wavelet decomposition was used and Symlets were used as the mother wavelet.

Three well established ML models were adopted in this research, namely, SVM, KNN, and LDA. Among the system design decisions was to incorporate mostly non-parametric models. This is meant to avoid any pre-assumptions about the data distribution. The settings of the three models was to their default in the EEG Lab library. We train and test each of the adopted models on a subject-by-subject basis, i.e., we train the models using samples from subject X then test the trained models using samples from the same subject. This was repeated with each of the other participants. In the next section, we report the results from this procedure.

\section{Results and Discussion}
The simulations conducted during the training and testing stages of this study were done on a Windows machine with an Intel core-i5 processor and 8GB of RAM. As stated in the previous section, three ML models were adopted in this research which are the LDA, SVM, and KNN classifiers. All these classifiers were trained to recognize stress/non-stress states from EEG signals. The results shown in this section indicate the performance of those classifiers during the testing phase. 

Conventional performance metrics, such as the accuracy and the F1-score, are usually used to reflect the performance of a model. However, this study is concerned with the “comparative” performance, that is the detected stress with and without VR-based attention training. Hence, in this section, we report absolute figures for the number of samples recognized as stress samples without receiving attention training prior to the exposure to stress and with receiving attention training prior to the exposure to stress. 

As stated in the previous section, the testing procedure of the classifiers was carried out on the EEG signals acquired from three participants, after being trained on ten participants. The three participants included in the testing procedure were among the ten participants of the training procedure. Hence, we already have EEG samples representing their stress and relaxed states. Indeed, these samples were acquired with the other samples that were used for training except that no overlap, of course, was allowed between training and testing samples. We refer to these samples as Case A in the analysis below. To complete the testing procedure, we included the EEG samples acquired from those three participants, while being exposed to the stressor, after receiving the VR-based AT. We refer to these samples as Case B in the analysis below. 

Fifty samples from Case A were considered during testing, while three temporal windows of fifty samples each, were considered from Case B. The three temporal windows were chosen to span the beginning, the middle, and the end of the EEG acquisition time. By this choice of the position of the temporal windows, we mean to show the consistency of the ML models performance across time. The second column of tables 1-3 indicate the results on Case A samples, while columns from 3-5 show the results on Case B samples with its three temporal windows for each of the three participants included in the testing phase.

As can be noticed from Table 1, which shows the results for the SVM classifier, the total number of Case A samples that were recognized as stress for participants 1, 2, and 3 is 127. This number has significantly been reduced to 85, 83, and 77 (depending on the position of the temporal window) when the subjects get VR-based AT before the exposure to the MAE stressor. By taking the average of the three temporal windows, the average number of samples recognized as stress samples for participants 1, 2, and 3 are 36.3, 18.6, and 26.6, respectively. The total number of samples for the three participants is approximately 82 samples ($36.3+18.6+26.6$) samples out of 150 Case B samples, while the total is 127 for Case A samples. This translates to 30\% reduction in the number of stress samples. Similarly, the percentage reduction for the KNN classifier is 59\%, and the percenatge reduction for the LDA classifier is 52\%.

% \usepackage{multirow}
\begin{table}[]
\caption{The number of samples that has been identified as stress using the SVM classifier. The second column shows Case A samples while the last three columns show three different temporal windows for Case B samples. Please see text for more details.}
\begin{center}
\begin{tabular}{|c|c|ccc|}
\hline
                                        &                                & \multicolumn{3}{c|}{\textbf{After Attention Training based VR module}}                                                      \\ \cline{3-5} 
\multirow{-2}{*}{\textbf{Participants}} & \multirow{-2}{*}{\textbf{MAE}} & \multicolumn{1}{c|}{\textit{1st 50 samples}}    & \multicolumn{1}{c|}{\textit{2nd 50 samples}}    & \textit{3rd 50 samples} \\ \hline
1                                       & 49                             & \multicolumn{1}{c|}{35}                         & \multicolumn{1}{c|}{34}                         & 40                      \\ \hline
\rowcolor[HTML]{F2F2F2} 
2                                       & 38                             & \multicolumn{1}{c|}{\cellcolor[HTML]{F2F2F2}24} & \multicolumn{1}{c|}{\cellcolor[HTML]{F2F2F2}23} & 9                       \\ \hline
3                                       & 40                             & \multicolumn{1}{c|}{26}                         & \multicolumn{1}{c|}{26}                         & 28                      \\ \hline
\rowcolor[HTML]{F2F2F2} 
Total                                   & 127                            & \multicolumn{1}{c|}{\cellcolor[HTML]{F2F2F2}85} & \multicolumn{1}{c|}{\cellcolor[HTML]{F2F2F2}83} & 77                      \\ \hline
\end{tabular}
\end{center}
%\end{table}

%\vspace{-5mm}
%========================
%\begin{table}[]
\caption{The number of samples that has been identified as stress using the KNN classifier. The second column shows Case A samples while the last three columns show three different temporal windows for Case B samples. Please see text for more details.}
\begin{center}  
\begin{tabular}{|c|c|ccc|}
\hline
                                        &                                & \multicolumn{3}{c|}{\textbf{After Attention Training based VR module}}                                                      \\ \cline{3-5} 
\multirow{-2}{*}{\textbf{Participants}} & \multirow{-2}{*}{\textbf{MAE}} & \multicolumn{1}{c|}{\textit{1st 50 samples}}    & \multicolumn{1}{c|}{\textit{2nd 50 samples}}    & \textit{3rd 50 samples} \\ \hline
1                                       & 44                             & \multicolumn{1}{c|}{6}                          & \multicolumn{1}{c|}{7}                          & 8                       \\ \hline
\rowcolor[HTML]{F2F2F2} 
2                                       & 36                             & \multicolumn{1}{c|}{\cellcolor[HTML]{F2F2F2}1}  & \multicolumn{1}{c|}{\cellcolor[HTML]{F2F2F2}4}  & 12                      \\ \hline
3                                       & 39                             & \multicolumn{1}{c|}{12}                         & \multicolumn{1}{c|}{18}                         & 23                      \\ \hline
\rowcolor[HTML]{F2F2F2} 
Total                                   & 119                            & \multicolumn{1}{c|}{\cellcolor[HTML]{F2F2F2}19} & \multicolumn{1}{c|}{\cellcolor[HTML]{F2F2F2}29} & 43                      \\ \hline
\end{tabular}
\end{center}
%\end{table}

%\vspace{-5mm}
%========================
%\begin{table}[]
\caption{The number of samples that has been identified as stress using the LDA classifier. The second column shows Case A samples while the last three columns show three different temporal windows for Case B samples. Please see text for more details.}
\begin{center}    
\begin{tabular}{|c|c|ccc|}
\hline
                                        &                                & \multicolumn{3}{c|}{\textbf{After Attention Training based VR module}}                                                      \\ \cline{3-5} 
\multirow{-2}{*}{\textbf{Participants}} & \multirow{-2}{*}{\textbf{MAE}} & \multicolumn{1}{c|}{\textit{1st 50 samples}}    & \multicolumn{1}{c|}{\textit{2nd 50 samples}}    & \textit{3rd 50 samples} \\ \hline
1                                       & 42                             & \multicolumn{1}{c|}{20}                         & \multicolumn{1}{c|}{22}                         & 21                      \\ \hline
\rowcolor[HTML]{F2F2F2} 
2                                       & 37                             & \multicolumn{1}{c|}{\cellcolor[HTML]{F2F2F2}0}  & \multicolumn{1}{c|}{\cellcolor[HTML]{F2F2F2}7}  & 6                       \\ \hline
3                                       & 44                             & \multicolumn{1}{c|}{23}                         & \multicolumn{1}{c|}{13}                         & 22                      \\ \hline
\rowcolor[HTML]{F2F2F2} 
Total                                   & 123                            & \multicolumn{1}{c|}{\cellcolor[HTML]{F2F2F2}43} & \multicolumn{1}{c|}{\cellcolor[HTML]{F2F2F2}42} & 49                      \\ \hline
\end{tabular}
\end{center}
\end{table}
%========================

\section{Conclusion}
In this research, we investigated the applicability of attention exercises consumed through VR headsets in stress management contexts. Although the literature on tech-based tools for stress management and the diverse applications of VR is quite immense, we argue that the literature on the capacity virtual attention training in that specific context still lacks more investigation, which is the inspiration behind this research. Towards addressing this gap, we presented a complete study design in addition to the implementation details. We started by training diverse ML classifiers on stress detection using ten undergraduate engineering students. Then, we recruited three students for testing. We exposed them to mental arithmetic (MAE) stress inducing tasks after receiving VR-based attention training. Using multiple extracted features, namely, MFMD and wavelet features, and three ML models, namely, SVM, KNN, and LDA, we show that preceding stressors by VR-based training results in reducing the number of recognized MAE-induced stress samples in the recorded EEG signals by 30\%, 59\%, and 52\%, respectively. This study presents preliminary results. Hence, towards acquiring better insights on the feasibility of virtual attention training in this context, we stress the necessity of future research studies that involve a larger number of participants and different types of stressors.

\bibliographystyle{splncs04}
\bibliography{mybibliography}

%Author, F.: Article title. Journal \textbf{2}(5), 99--110 (2016)
%\end{thebibliography}
\end{document}